\documentstyle{article}
\begin{document}

\title
{
Determination of Exchange Parameters from Magnetic Susceptibility 
}

\author
{ 
Ken'ichi Takano \\
Toyota Technological Institute, Tenpaku-ku, Nagoya 468 \\
Kazuhiro Sano \\
Department of Physics, Faculty of Education, \\
Mie University, Tsu, Mie 514
}

\maketitle

\begin{abstract}
      We report a novel practical method to 
determine exchange parameters by using experimental 
susceptibility data in a relatively narrow temperature region 
and a low order high-temperature-expansion equation. 
      This method is applied to a square lattice and a CaV$_4$O$_9$-type 
lattice, and its accuracy is discussed. 
\end{abstract}

\vskip 0.5cm


\vskip 1cm


      When we describe magnetic properties of a matter 
by a Heisenberg model, we must determine exchange 
parameters in the model to fit the matter. 
      It is sometimes performed by the high temperature 
expansion (HTE) of the susceptibility of the Heisenberg model 
and by the comparison of it to the experimental data. 
      However meaningful experimental data are often 
restricted in a low temperature region because of, {\it e.g.}, the 
temperature dependence of the lattice constant or structure. 
      Further it becomes hard to calculate high order terms 
of the HTE because of striking increase of 
the number of diagrams if the model includes many exchange 
parameters. 
     In some of such cases, an ordinary method of the direct 
fitting may not estimate exchange parameters with 
sufficient accuracy. 

      In this article, we report a novel practical method to 
determine exchange parameters by using experimental 
susceptibility data in a relatively narrow temperature region 
and a low order HTE equation. 
      The essence of this method is that we construct an experimental 
formula for the experimental susceptibility data in a power series 
of 1/T and then compare the coefficients of this formula to those 
of the HTE equation derived from the assumed Heisenberg model. 


      We denote the experimental data for the magnetic susceptibility 
by $\chi^{EXP}(T)$ and the experimental formula in a power series 
of 1/T by $\chi(T)$. 
      To make expansion coefficients dimensionless 
we use $x=T_0/T$ as the expansion parameter, where $T_0$ is 
a constant temperature chosen arbitrarily. 
      Then $\chi(T)$ is written as 
\begin{eqnarray} 
\label{chi} 
     \chi(T) &=& \lim_{n \rightarrow \infty} \chi^{(n)}(T) , \nonumber \\
     \chi^{(n)}(T) &=& \frac{C}{T} \sum_{m=0}^{n} A_m x^m ,
\end{eqnarray}
where  $C$ is Curie constant determined by experiments.
      The zeroth expansion coefficient is set as $A_0$ = 1 and 
the others ($A_m$'s for $m \ge 1$) are fitting parameters 
to be determined. 

      Using the same coefficients $A_m$'s, we define the 
following quantity:
\begin{equation}
\label{phiFITd}
   \phi_m(x) = A_m + \sum_{l=1}^{\infty} A_{m+l} x^l 
\end{equation}
for each $m$.  
      In the high temperature limit, it reduces to
a coefficient as 
\begin{equation}
\label{phiFITh}
      \phi_m(0) = A_m .  
\end{equation}
     This quantity $\phi_m(x)$ is also defined by the recursion 
equation: 
\begin{eqnarray}
\label{phiFIT}
   \phi_0(x) &=& \frac{T}{C} \ \chi(T) , \nonumber \\ 
   \phi_m(x) &=& (\phi_{m-1}(x) - A_{m-1}) 
   \frac{1}{x}
\end{eqnarray}
for positive integer $m$.  

      The experimental data corresponding to $\phi_m(x)$ 
is similarly defined in the same recursive transformation:  
\begin{eqnarray}
\label{phiEXP}
   \phi^{EXP}_0(x) &=& \frac{T}{C} \ \chi^{EXP}(T) , \nonumber \\ 
   \phi^{EXP}_m(x) &=& (\phi^{EXP}_{m-1}(x) - A_{m-1}) 
   \frac{1}{x} .
\end{eqnarray}
      The quantity $\phi^{EXP}_m$ is actually determined 
only when $A_{m-1}$ is known. 
      It is also noticed that $\phi^{EXP}_m$ is defined only 
at experimental data points in $\{x\}$; 
      e.~g. it is not defined at very high temperatures 
($x \sim 0$). 
      $\phi_m$ must fit $\phi^{EXP}_m$, since $\chi$ is 
constructed to fit $\chi^{EXP}$ and $\phi_m$ is produced 
by the same transformation as $\phi^{EXP}_m$. 

      In the following, we inductively determine $\{A_m\}$ 
together with $\{\phi^{EXP}_m(x)\}$. 
      For $m$ = 0, $\phi^{EXP}_0(x)$ is defined by eq.~(\ref{phiEXP}) 
and $A_0 = 1$ is a definition. 
      Assuming that $\phi_{m-1}(x)$ and $A_{m-1}$ are known, 
we construct $\phi^{EXP}_m(x)$ and $A_m$. 
      $\phi^{EXP}_m(x)$ is simply given by eq.~(\ref{phiEXP}).  
      To determine $A_m$, we consider the high temperature 
limit ($x \rightarrow 0$) for $\phi^{EXP}_m(x)$ which 
corresponds to eq.~(\ref{phiFITh}) for $\phi_m(x)$. 
      Since $\phi^{EXP}_m(x)$ does not have data points 
in the vicinity of $x$=0, we extrapolate the graph of 
$\phi^{EXP}_m(x)$ to $x=0$ by using a polynomial fitting 
function: 
\begin{equation}
\label{fm}
   f_m(x) = \sum_{l=0}^{L} a_{ml} x^l, 
\end{equation}
where coefficients $a_{ml}$'s $(l=0, \cdots , L)$ are 
fitting parameters to be determined for each $m$. 
      The order $L$ of the polynomial is empirically chosen as 
8 to 10 and is inspected with some examples as mentioned later. 
      $A_m$ is then determined by $A_m$ $\simeq$ $f_m(0)$ = 
$a_{m0}$ corresponding to eq.~(\ref{phiFITh}).  
      Thus we have completed the inductive definition and 
have constructed an experimental HTE formula (1b). 


      On the other hand, we start from a Heisenberg Hamiltonian 
$H$ which is assumed to describe the magnetic properties of 
the material. 
     The assumed Hamiltonian includes a set of exchange 
parameters \{$J_i$\}. 
     For this Hamiltonian, we calculate the theoretical HTE 
formula 
\begin{equation}
\label{chiHTE}
   \chi^{HTE}(T) = \frac{C}{T} \sum_{m=0}^{\infty} F_m x^m 
\end{equation}
by the standard diagrammatic method.~\cite{Rushbrooke} 
     The expansion coefficients $F_m$'s are functions of the 
set of exchange parameters $\{J_i\}$. 

      Lastly we compare the theoretical HTE formula (\ref{chiHTE}) 
to the experimental formula (\ref{chi}). 
      If the assumed Hamiltonian exactly describes the material, 
both the coefficients must be the same. 
      Hence we can determine the exchange parameters 
by solving the set of equations 
\{$F_m(\{J_i\}) = A_m$\}. 
      In real cases, however, the model Hamiltonian may not 
perfectly describe the matter and the experimental data 
may include some errors. 
      We usually must be satisfied with a set $\{J_i\}$ which 
gives approximate matching between \{$F_m$\} and \{$A_m$\}. 


      We inspect the method mentioned above for the Heisenberg 
model in a simple square lattice. 
      Only an exchange parameter $J$ is between nearest neighbor 
spins. 
      As data for susceptibility, we use the result of Quantum 
Monte Carlo (QMC) simulation by Troyer et al.~\cite{Troyer} 
instead of data of a real experiment. 
      We choose the expansion parameter as $x = 1/T$ in the unit of 
$T_0 = 1$ and use 12 QMC data points in $0.05 \le T \le 1.5$. 
      The data is for $J=1$ and $C$ = 0.25. 

      Following the above procedure, we obtained coefficients 
of HTE as $A_1$ = $-0.994$, $A_2$ = $0.483$ and 
$A_3$ = $-0.135$ for $L$ = 8. 
      These values are little changed when the order $L$ of the fitting 
function changes from 6 to 10. 
      We show $\phi^{EXP}_m(x)$ along with $f_m(x)$ for 
$m$ = 1 to 3 in Fig.~\ref{fig:phi_m}. 

      The corresponding theoretical coefficients are calculated 
as $F_1 = -J$, $F_2 = J^2/2$ and $F_3 = -J^3/6$. 
      In this one-parameter case, equation $F_1(J) = A_1$ 
is sufficient to determine $J$. 
      The solution is $J = 0.994$, which is close to 
the exact value $J=1$. 
      We estimate the accuracy of $A_m$ by the deviation 
$d_m = 2 | (A_m - F_m(1))/(A_m + F_m(1)) |$. 
      It is given as $d_m$ = 0.006, 0.017 and 0.070 for 
$m$ = 1, 2 and 3, respectively. 
      This result, i.~e. this method, is fairly accurate 
despite the small number of the data points. 
\begin{figure}
\caption{Transformed QMC data $\phi^{EXP}_m(x)$ (circle) 
and fitting function $f_m(x)$ (line) with $m$ = 1 to 3 
for a square lattice.} 
\label{fig:phi_m}
\end{figure}

      We also carried out the same analysis by restricting the 
temperature region as $0.05 \le T \le 0.9$, which includes 
only 9 data points. 
      This region is below the maximum point of $\chi^{EXP}(T)$ 
and is rather a low temperature region. 
      The present method gives $A_1$ = $-1.019$, 
$A_2$ = $0.541$ and $A_3$ = $-0.185$ for $L$ = 8. 
      The corresponding deviations are $d_m$ = 0.019, 0.079 
and 0.327 for $m$ = 1, 2 and 3, respectively. 
      This estimation is practically sufficient to determine 
$J$ from $A_1$. 


      Next we apply our method to a Heisenberg model on 
a lattice which is a simplified two-dimensional model 
used for CaV$_4$O$_9$. 
      This model includes two exchange parameters, 
$J_0$ for the plaquette link and $J_1$ for the dimer link. 
      Theoretical coefficients are obtained as 
$F_1(J_0, J_1) = -(2J_0 + J_1)/4$ and 
$F_2(J_0, J_1) = (4J_0 - J_1) J_1/16$. 
      Instead of experimental data, we use QMC data~\cite{Troyer} 
and apply our method to determine $J_0$ and $J_1$. 
      We used 15 QMC data points in $0.075 \le T \le 1.5$ and 
the data are for $C$ = 0.25 in the unit of $T_0$ = 1. 
      We examine two typical cases of $J_0/J_1$ =1.0
and $J_0/J_1$ = 0.5. 
      The system has no spin gap in the former case, while it has 
in the latter.~\cite{Troyer}

      For the data of $(J_0, J_1)$ = (1, 1), our method produces 
coefficients, $A_1 = -0.750$ and $A_2 =  0.166$. 
      The fitting was done by the 8th order polynomial ($L=8$), $f_m$'s. 
      The corresponding exact values are $F_1 = -0.750$ and $F_2 = 0.188$, 
and the deviations are $d_1 = 0.00$ and $d_2 = 0.12$. 
     The complete matching between $A_1$ and $F_1$ 
over 3 digits are occasional, since the last digit changes 
if we change $L$ from 8 to 9 or 10. 
      We calculated the exchange parameters as 
$(J_0, J_1)$ = (0.83, 1.34) or (1.17, 0.66) by solving  
$F_1(J_0, J_1)$ = $A_1$ and $F_2(J_0, J_1)$ = $A_2$. 

      Similarly we applied our method for the data of 
$(J_0, J_1)$ = (0.5, 1). 
      We obtained HTE coefficients as $A_1 = -0.500$ and $A_2 =  0.052$. 
      The corresponding exact values are $F_1 = -0.500$ and 
$F_2 = 0.063$, and the deviations are $d_1 = 0.00$ and $d_2 = 0.19$. 
      The obtained exchange parameters are $(J_0, J_1)$ = (0.46, 1.08). 

      The estimation gives better result for a square lattice than 
for a CaV$_4$O$_9$ lattice. 
      A reason may be small HTE coefficients included in the latter lattice. 
      Such small coefficients seems to appear, 
when the coordination number is small so that the spin 
fluctuation is strong. 
      Large fluctuation is also a cause of the spin-gap formation. 
      The present method is applicable to gapful systems as well as 
gapless systems as long as we only need coefficients of low orders. 


      We thank K. Kubo for useful discussion on high temperature 
expansion of Heisenberg models.  
      One of us (K. T.) carried out this work partially in Department 
of Physics of Nagoya University as a Guest Associate Professor.  
      This work is partially supported by the Grant-in-Aid for 
Scientific Research from the Ministry of Education, Science 
and Culture, Japan.  
      The computation in this work was carried out partially by 
using facilities of the Supercomputer Center, Institute 
for Solid State Physics, University of Tokyo.  


\end{document}